# Water Mediated Electron Attachment to Nucleobases: Surface-bound *versus* Bulk Solvated Electrons

Madhubani Mukherjee[#], Divya Tripathi[#], and Achintya Kumar Dutta[*]

*Department of Chemistry, Indian Institute of Technology Bombay, Powai, Mumbai 400076, India*

We have presented a mechanism for electron attachment to solvated nucleobases using accurate wave-function based QM/MM simulations and uracil as a test case. The initial electron attached state is found to be localized on the bulk water, and this water-bound state acts as a doorway to the formation of the final nucleobase bound state. The electron transfer from water to uracil takes place due to the mixing of electronic and nuclear degrees of freedom. The water molecules around the uracil stabilize the uracil bound anion by creating an extensive hydrogen-bonding network and accelerate the rate of electron attachment to uracil. The complete transfer of the electron from water to the uracil occurs in a picosecond time scale, which is consistent with the experimentally observed rate of reduction of nucleobases in the presence of water. The degree of solvation of the aqueous electron can lead to a difference in the initial stabilization of the uracil bound anion. However, the anions formed due to the attachment of both surface-bound and bulk solvated electrons behave similarly to each other at a longer time scale.

[#] *The authors contributed equally*

[*]*achintya@chem.iitb.ac.in*

**A. Introduction:**

Radiation damage to genetic materials is one of the active fields in chemical research owing to its implications both in the cause and cure of cancer. High energy radiation can structurally alter DNA leading to base release, base damage, and strand break[1]. The creation of many of these local damage sites in the near vicinity leads to the formation of multiple damage sites (MDS)[2]. The formation of MDS can cause loss of genetic information, and they are often irreparable by the enzymatic repair process. The cause of such damages is, for a long time, attributed to ionization and excitation processes created by high energy radiation. However, Sanche and co-workers' detailed experimental results in recent times have suggested a critical role played by the low-energy secondary electrons in the radiation damage process[3,4]. The secondary electrons are generated by the action of the ionizing radiation on the water surrounding the genetic materials. They can get attached to the base, sugar, and phosphate group of the genetic materials causing structural distortion, ultimately leading the single and double-strand breaks by the cleavage of the phosphate-sugar C-O bond or base release by N1-C1 glycosidic bond cleavage[5]. The ionizing radiation acting on water initially produces pre-solvated electron ($e_{pre}^-$) in the water conduction band with a lifetime of 100 to 500 fs[6]. The surrounding water molecules progressively orient themselves along the ($e_{pre}^-$) due to the interaction of the charge with the induced or the permanent dipole moment of the water molecules and lead to partial and full solvation of the electron. The partially solvated electron can exist in the surface ($e_{aq}^-(s)$) of the water for a very short time and is transformed into bulk-solvated electron($e_{aq}^-(b)$) within a picosecond time scale. The presence of the aqueous environment is known to accelerate the radiation damage of genetic material[7]. Therefore, a detailed understanding of the mechanism of electron attachment to solvated nucleobases is of fundamental importance. The last few years have seen a plethora of experimental results[8,9] in the form of ultrafast transient spectroscopy, which can provide guidelines for understanding the role of water in accelerating the rate of electron attachment to nucleobases. The experiments have suggested the role of the excited state of the anion in the electron attachment process[9]. However, the exact mechanism is not yet understood.

Theoretical calculations can be a great aid in understanding the radiation damage pathways of genetic materials. Consequently, a large number of theoretical studies[10–15] are reported in the literature starting from DFT, MP2, CASPT2 to the state of the art coupled-cluster method. The environmental effect in these studies is mostly taken care of by the so-called polarizable continuum model (PCM)[16,17]. Simons and co-workers[18] have recently demonstrated that the PCM based method fails to give an accurate description of the solvated electron. Although some limited number of QMMM based explicit solvation studies are available for the electron attachment to nucleobases, the QM method used in those studies is DFT. Now, the standard DFT methods are particularly unsuitable for anions due to the self-interaction problem[19]. Moreover, the simulation of electron attachment to solvated nucleobases would require accurate computation of both ground and excited state of the anion, which is difficult to achieve in DFT (or its time-dependent analog) based method. The wave-function based equation of motion coupled cluster(EOM-CC) approach[20,21] can provide a balanced description of the ground and the excited state of the anion. However, the computational cost associated with coupled cluster based methods makes it difficult to perform the EOM-CC calculations beyond small molecules. This is particularly problematic for explicit solvation studies, where averaging over a large number of snapshots is required to accurately sample the solvent distribution. In recent times development of domain-based pair natural orbital[22,23,32–35,24–31] equation of motion coupled cluster approach[28,36] has made it possible to calculate the electron affinity of multiple electron attached states at a fraction of the computational cost of a canonical calculation and with a systematic controllable accuracy. In this paper, we propose a detailed mechanism for the first time from the high-level EOM-CCSD based QMMM calculations, which explains the experimentally observed rate of electron attachment to solvated nucleobases in water.

## B. Computational Details:

The geometry of the micro-solvated uracil has been optimized using the RI-MP2/aug-cc-pVDZ level of theory. The adiabatic potential energy surface corresponding to the linear transit from dipole-bound to valence-bound geometry has been calculated using EA-EOM-DLPNO-CCSD[28] method with NORMALPNO settings. The intermediate geometries were obtained by varying the linear transition parameter $\lambda$ from 0 to 1 in the following formula

$R = (1-\lambda) R_{DB} + \lambda R_{VB}$

where R is the geometrical parameter (bond length, bond angle and dihedral angle) for the intermediate geometry, $R_{DB}$ is the parameter for the dipole-bound geometry, and $R_{VB}$ is the parameter for the valence-bound anion geometry. The coupling elements between two diabatic states are obtained by fitting a simple avoided crossing model potential[37] defined as follows

$$V = \begin{pmatrix} V_1 & W \\ W & V_2 \end{pmatrix}$$

Where the diagonal elements are chosen to be harmonic potential in the coordinate $\lambda$ with the form

$$V_i = \frac{1}{2}\omega_i \left(\lambda - \lambda_i^0\right)^2 + v_i^0$$

and the off-diagonal element is assumed to be constant. We have calculated the rate of transition of an electron from dipole-bound to valence-bound state using the Marcus theory[38].

$$k = \frac{2\pi}{\hbar}|W|^2 \sqrt{\frac{1}{4\pi k_B T \lambda_R}} e^{-\frac{\left(\lambda_R + \Delta G^0\right)^2}{4\lambda k_B T}}$$

Where $\Delta G^0$ is the free energy change between the valence and dipole-bound state ($E_{VB}-E_{DB}$, ignoring the entropy contribution), $\lambda_R$ is the reorganization energy.

The classical molecular dynamics simulations on the uracil are performed in NAMD[39] using a CHARMM compatible forcefield. The TIP3P model has been used for the water. Uracil was solvated with 2460 water molecules (TIP3P) in a cubic box of 40 Å, and minimization was performed keeping the uracil fixed and using the non-bonded cutoff of 12 Å. The minimized structure was heated up to 300 K using Langevin Thermostat. After the heating was complete, the system was equilibrated for 500 ps using the time step for integration as 2 fs. Periodic boundary condition and particle-mesh Ewald technique (PME) was applied during the equilibration run. Subsequently, a 10 ns Production run was carried out in constant pressure (1 bar) - constant temperature (300 K) condition with Nosé-Hoover Langevin piston pressure control. From the Production run, 10 snapshots were generated in a regular interval of 1 ns.

The snapshot after the equilibration run and the snapshot after 1ns of production run were taken for mixed quantum-classical QM/MM simulation using ORCA-NAMD QM/MM interface[40]. The uracil is treated as the QM region at BP86 and def2-SVP basis set, and waters are treated at the MM level using the same forcefield as that of the classical MD simulations. The timestep for integration was taken as 0.5 fs, and the QM/MM simulations were run for 2ps in both the cases. In each case, 4000 snapshots were generated from the QM/MM trajectory in an interval of 0.5 fs. The single point QM/MM calculation on each snapshot was performed using ORCA[41] with uracil treated using EA-EOM-DLPNO-CCSD method[28] with NORMALPNO setting, and the TIP3P model has been used for the water. The aug-cc-pVDZ basis sets with additional 5s5p4d diffuse functions added on the carbonyl oxygen of uracil have been used for all the calculations. The potential energy surface of micro-solvated uracil and rates of electron transfer in bulk water are calculated using aug-cc-pVTZ basis sets with additional 5s5p4d diffuse functions. Periodic boundary conditions were turned off for the single-point EA-EOM-DLPNO-CCSD calculations to avoid artifacts due to the diffuse basis functions. The snapshots from the QM/MM simulations using geometry from the equilibration run lead to partially solvated electrons, and snapshots from the QM/MM simulation on the geometries from the production run leads to bulk solvated electrons. Multiple trajectories were considered for the analysis. More details on standardizing the QM/MM protocol has been provided on the supporting information.

## C. Results and Discussions:

### C.1 Microhydration

The monohydrated uracil provides a simple model system to investigate the role of water in the process of electron attachment to nucleobases. The electron attachment to monohydrated uracil at the neutral geometry leads to the formation of the dipole-bound state, which is vertically bound with a vertical detachment energy value of 0.069 eV. The additional electron is delocalized away from the nuclear framework (see the natural orbitals in Figure 1(a)). Consequently, the formation of the dipole-bound anion does not lead to any distortion from the neutral geometry. There also exits a valence-bound anionic state, which is adiabatically bound with AEA of 0.142 eV. The valence bound state shows vertical detachment energy of 0.859 eV. The additional electron in the valence-bound state is localized on the nuclear framework, and the formation of the valence-bound anionic state leads to geometric distortion of the uracil. The transition between the two anionic states is very feebly optically allowed with an oscillator strength of only 0.004. However, the molecular vibrations can result in the inter-conversion between the valence and the dipole-bound state of the anion. To understand that, we have plotted the adiabatic potential energy surface (PES) of the ground and the first excited state of the anion along with a linear transit from the dipole-bound to valence-bound geometry. The dipole and valence-bound state correspond to two different minima in the ground state adiabatic potential energy surface, and there exists an avoided crossing between the ground and the first excited state of the anion. The nature of the electronic states changes very rapidly near the avoided crossing, which increases the magnitude of the derivative coupling term and leads to the breakdown of the Born-Oppenheimer approximation. In this kind of situation, it is more advantageous to treat the two states in terms of diabatic potential energy surfaces, which crosses each other along the path connecting the two minima, and the dominant part of the coupling term is shifted to the electronic Hamiltonian.

The valence and dipole-bound nature of the anionic states give us an obvious choice for the diabatic basis[42]. We have generated the two diabatic states by fitting a harmonic potential to the valence-bound and dipole-bound part of the ground state adiabatic PES of the mono-hydrated uracil anion. At the dipole-bound geometry ($\lambda=0$) only one bound anionic state exists, and that is dipole-bound in nature. The second bound anionic state appears while transiting from the dipole-bound to the valence-bound geometry ( around $\lambda=0.4$). This second bound anionic state is valence-bound and is an excited state of the dipole-bound anionic ground state. Eventually, the valence-bound state becomes the ground state as the monohydrated uracil anion evolves towards the minimum energy structure. This is similar to the 'doorway mechanism' observed in electron attachment to the gas phase nucleobases, where the initial electron attachment leads to a vertically bound dipole-bound state. Subsequently, the dipole-bound state gets transformed into an adiabatically bound valence-bound state. It can be seen from Figure S1 that the valence bound state in isolated uracil is not adiabatically bound. Therefore, it is the presence of the water molecules which makes the doorway mechanism[11,42–44] operative in the microsolvated uracil. The rate of transition from the dipole-bound to valence-bound state is in the microsolvated uracil is $1.2\times10^{11}$ sec$^{-1}$, which is three-fold higher than that observed in the isolated uracil. Therefore, even a single water molecule can significantly increase the rate of transfer of electron. However, the rate is still one order of magnitude lower than the experimentally observed rates[8] for the reduction of nucleobases in bulk water. The electron attachment process to uracil in bulk water can significantly differ from that observed in the simple micro-solvated model. To understand that, we have simulated the electron attachment dynamics of both $e_{aq}^-(s)$ and ($e_{aq}^-(b)$) in bulk-water using QM/MM molecular dynamics simulation.

C.2. *Bulk solvation:*

Simons and co-workers[18] have recently shown that the dipole bound electron attached states cannot survive in bulk water. However, the water-bound anionic states can also act as the doorway for electron attachment to solvated nucleobases. Figure 2 presents the natural orbital corresponding to the time evolution of the anionic state formed due to the attachment of a partially solvated electron. The initial attachment (0 fs) of the partially solvated electron to uracil leads to the formation of two vertically bound anionic states. The ground state has a VDE of 0.089 eV, and the additional electron is localized on the surface of the water. The excited state is also surface-bound and has a VDE of 0.023 eV. The natural orbitals corresponding to the additional electron in both the anionic states are away from the nuclear framework of the nucleobase, similar to the dipole-bound anionic state in micro-solvated uracil. Bound anionic state localized on the nucleobase first appears around 5 fs and has a VDE of 0.036 eV. The uracil-bound state initially appears as an excited state of the anion, and the corresponding natural orbital appears similar to the valence-bound state of mono-hydrated uracil anion. The ground state of the anion is still delocalized over the water surface with VDE of 0.097 eV. The transition between the water-bound state and uracil-bound state is very weakly allowed (oscillator strength 0.003) as in the case of the dipole- and the valence-bound states of monohydrated uracil anion. The uracil bound state almost immediately transforms (at 5.5fs ) into the ground state with a VDE of 0.249 eV, and the first excited state of the anion is localized at the water surface. Therefore, the water-bound states act as a doorway for the formation of the final uracil bound state. The plot of detachment energy with respect to time shows the sign of avoided crossing similar to that observed in monohydrated uracil, indicating the mixing of electronic and nuclear degrees of freedom. The presence of large numbers of water molecules makes it difficult to use any sophisticated diabatization procedure. We have plotted the time evolution of the detachment energy for the lowest uracil-bound and water-bound states (Figure 3 (a)) by visual inspections of the natural orbitals corresponding to the electron attached states. It can be seen that the detachment energy of the water-bound state increases very slowly. In contrast, the detachment energy corresponding to the uracil-bound state shows a steep decrease, and that leads to a flipping of the energy ordering of the two states.

The vertical attachment of $e_{aq}^-(b)$ to uracil in bulk water leads to the formation of multiple stable bound anionic states as opposed to only two bound states observed for partially solvated electrons. All of them are localized on the interior of bulk water (see Figure S9), and the anionic ground state has vertical detachment energy (VDE) of 0.739 eV, which is much larger than that observed for the partially solvated electron. The first four excited states of the anion are also bound, and corresponding excitation energies are 0.338 eV, 0.513 eV, 0.627 eV, and 0.721 eV, respectively. Bound electron attached states which were localized on the uracil start to appear within 3 fs (Figure 4). Initially, the uracil-localized state appears as an excited state and is very weakly bound with vertical detachment energy of 0.146 eV. The anionic ground state is still localized on the water, and the transition from water-bound anionic state to uracil-bound anionic state is very weakly optically allowed with an oscillator strength of 0.002, similar to that observed for the attachment of prehydrated electron. Within 6 fs, the uracil-bound state becomes the ground state with VDE of 0.988 eV. The water-bound state appears as the first excited state at 6 fs with an excitation energy of 0.121 eV. The plot of the detachment energy with respect to time shows avoided crossing similar to the partially solvated uracil anion (Figure S2). However, the presence of multiple avoided crossing indicates the presence of multiple electron trapping sites in the water. The electron can hop from the initial water-bound state to the final uracil bound state through these intermediate trapping sites. The detachment energy corresponding to the diabatic uracil-bound anionic state formed due to the attachment of bulk solvated electron increases much steeply than that in partially solvated one. The detachment energy of bulk-solvated uracil increases up to 500 fs around, after which they reach their equilibrium value of 4-5 eV, which is larger than the detachment energy (3.3 eV) of an electron in bulk water. The fluctuations in the detachment energy are more prominent for the partially solvated electron that reaches its equilibrium value around 1000 to 1500 fs (See Figure 3(c) ), which is consistent with the experimental time scale of conversion of the partially solvated electron to bulk solvated electron. Although the quantitative trends in the detachment energy are initially different for partially solvated and bulk solvated uracil anion, in the longer time scale (above 1.5 ps) both lead to similar detachment energy values.

The very high detachment energy of nucleobase in bulk water can be rationalized in terms of the local structure of the water distribution around the uracil anion. Figure 3(d) and 3(e) present the radial distribution function plot for O of uracil and H of water for the attachment of $e_{aq}^-(b)$ to uracil. The radial distribution function of both the uracil oxygen with the hydrogens of water shows a very sharp peak around 2Å for the anionic nucleobase. The peak appears in addition to the standard hydrogen bonding peak at 3Å present for solvated neutral nucleobases. It indicates the formation of a short-range local structure around the bulk solvated uracil as it evolves from the neutral to anionic state. The local structure of water preferentially stabilizes the uracil anion through its hydrogen bonding network and leads to a high detachment energy in solvated nucleobase anions.

Based on our theoretical calculations, we have proposed a two-step mechanism of reduction of nucleobases in water.

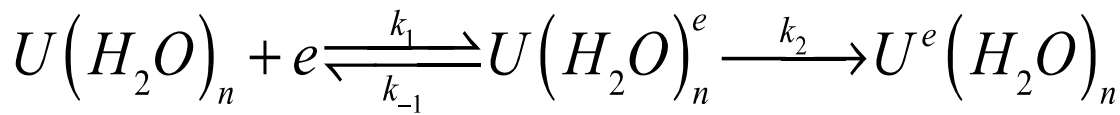

$$U(H_2O)_n + e \underset{k_{-1}}{\overset{k_1}{\rightleftharpoons}} U(H_2O)_n^e \xrightarrow{k_2} U^e(H_2O)_n$$

In the first step, the pre-solvated electron reacts with the solvated uracil, and the resulting anion gets solvated to give a uracil water anionic complex, where the extra electron is localized on water. In the second step, the electron transfer happens from the water to the nucleobase in the anionic complex. The solvation of pre-solvated electron takes place in the picosecond timescale[45]. To explain the experimentally observed rate[8] of reduction of nucleobases (0.6 ×$10^{12}$ M sec$^{-1}$ to 5 ×$10^{12}$ M sec$^{-1}$) using the above-mentioned mechanism, the electron transfer from water to uracil needs to happen at the picosecond time scale. The calculated rate of electron transfer from water-bound state to uracil bound state in bulk water is 2.1×$10^{12}$ sec$^{-1}$, which is consistent with the experimentally observed rate.

## D. Conclusions:

In this paper, we have presented a mechanism for electron attachment to solvated nucleobases from high-level QMMM calculations. It can be seen that the initial electron attachment to solvated nucleobases leads to the formation of the water-bound states, which acts as a doorway to the formation of the nucleobase bound state. The transfer of electrons from the initial water-bound to the final nucleobase bound state happens due to a mixing of the ground and excited states of the anion through vibronic coupling. The electron transfer happens at a picosecond timescale which is consistent with the experimentally measured rate of reduction of nucleobases in bulk water. The water molecules around the nucleobase anion play a crucial role in stabilizing the nucleobase bound anionic state through an extensive hydrogen-bonding network and accelerate the rate of electron transfer from the initial water bound to the final nucleobase bound state. The anionic states formed by the attachment of the partially solvated and the bulk solvated electron differ initially but leads to a similar nucleobase bound states on a longer time scale.

The mechanism of electron attachment presented in this paper can lead to new insights into the secondary radiation damage pathways of genetic material. Therefore, it will be interesting to extend this study to the electron attachment of other solvated nucleobases and larger models of genetic material. Work is in progress towards that direction[46].

## Supporting Information

The Supporting Information is available. The diabatic, adiabatic potential energy surface and the rate of transfer of electron in isolated uracil, the time evolution of the diabatic and adiabatic surface of the uracil anion in bulk water, the natural orbitals corresponding to all the bound anionic states, the details of the calculation of electron transfer rate in bulk water, and data on standardizing the QM/MM methodology are provided in the supporting information.

## Acknowledgment

The authors acknowledge the support from the IIT Bombay, IIT Bombay Seed Grant project, DST-Inspire Faculty Fellowship and SERB project no CRG/2018/001549 for financial support, IIT Bombay super-computational facility, Department of Chemistry Computational Facility and C-DAC Supercomputing resources (PARAM Yuva-II) for computational time.

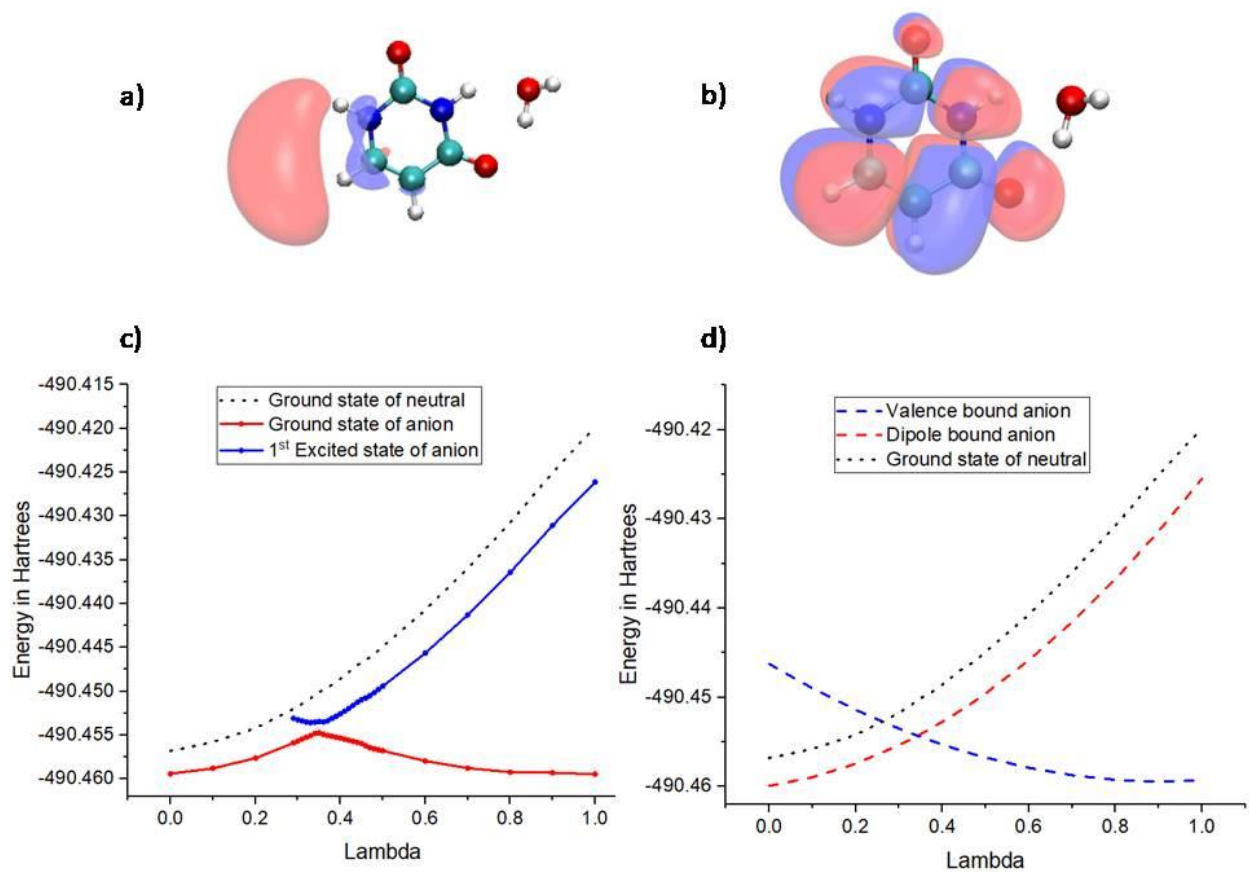

*Figure 1: Uracil monohydrate anion (a) Natural orbital corresponding to the dipole-bound state. (b) Natural orbital corresponding to the valence-bound state. (c) Adibatic surface corresponding to the ground and first excited state (d) Diabatic dipole and valence-bound state*

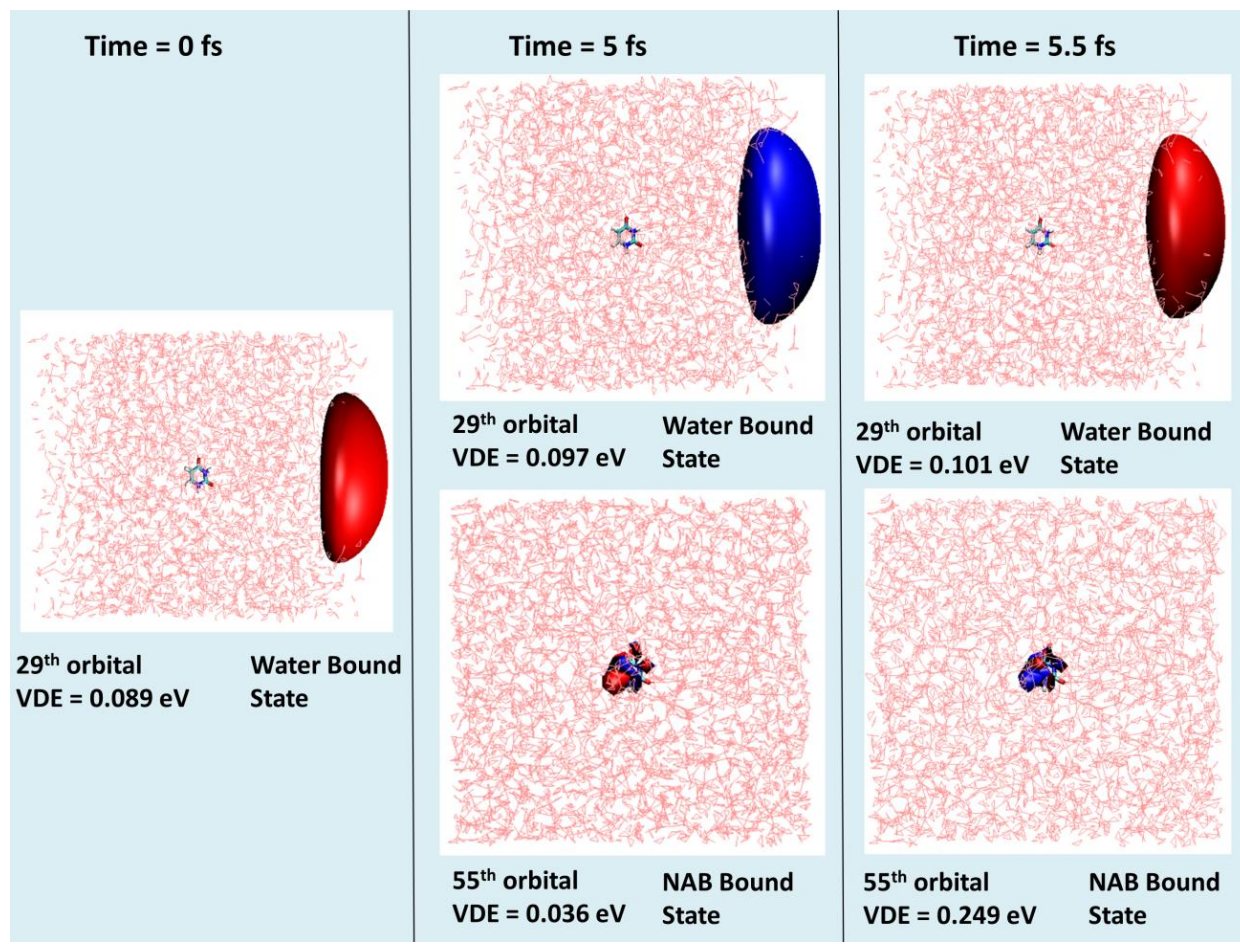

*Figure 2: EA-EOM-DLPNO-CCSD natural orbitals depicting time evalution of anionic state formed by attachment of partically solvated electron to uracil*

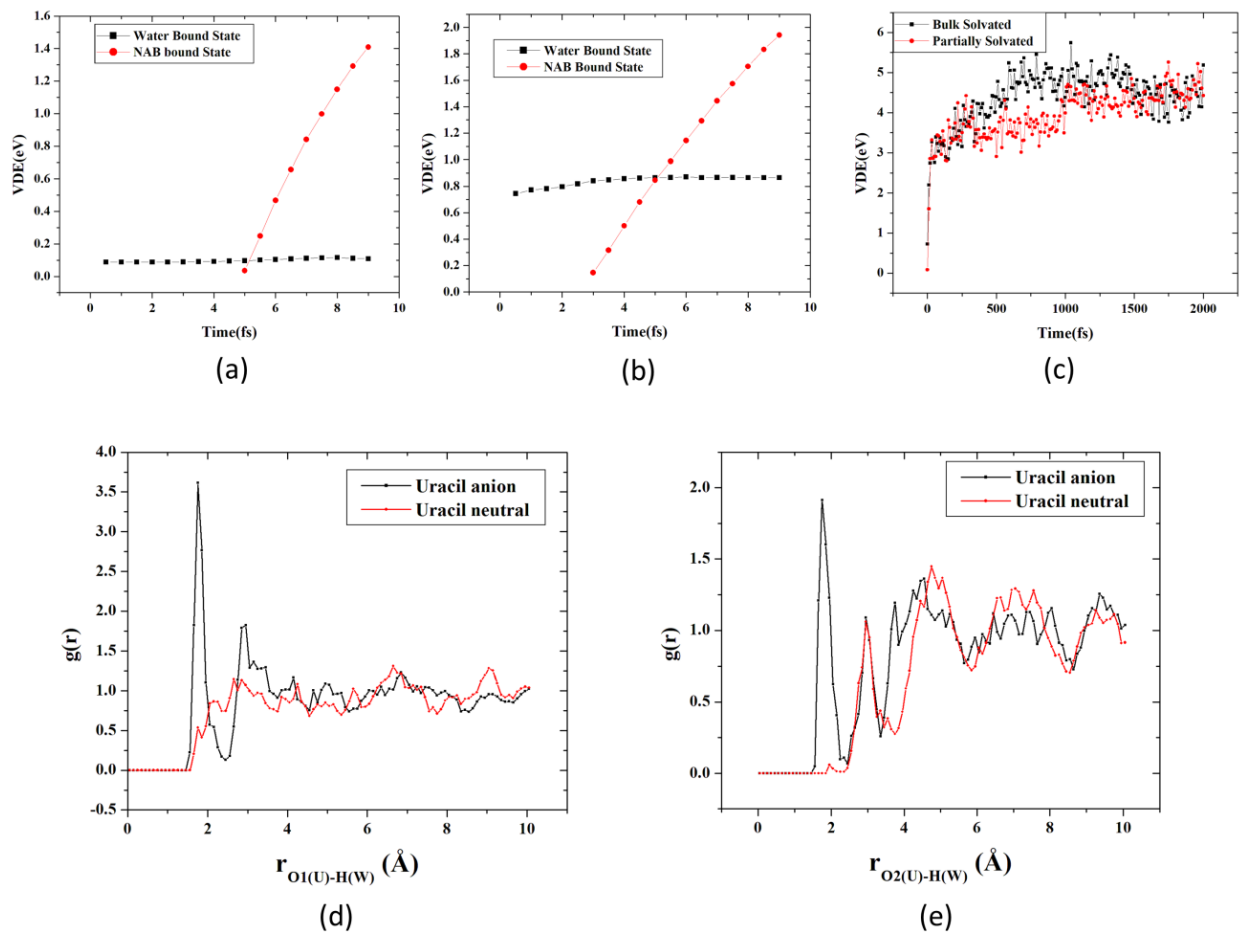

*Figure 3: The upper panel depicts time evalultion of detachment energy of the valence and dipole states for the attachment of (a) partially solvated (b) bulk solvated (c) both on a longer time scale. The lower panel depicts the radial distribution function (RDF) plot for (d) O1 of uracil and H of water (e) O2 of uracil and H of water*

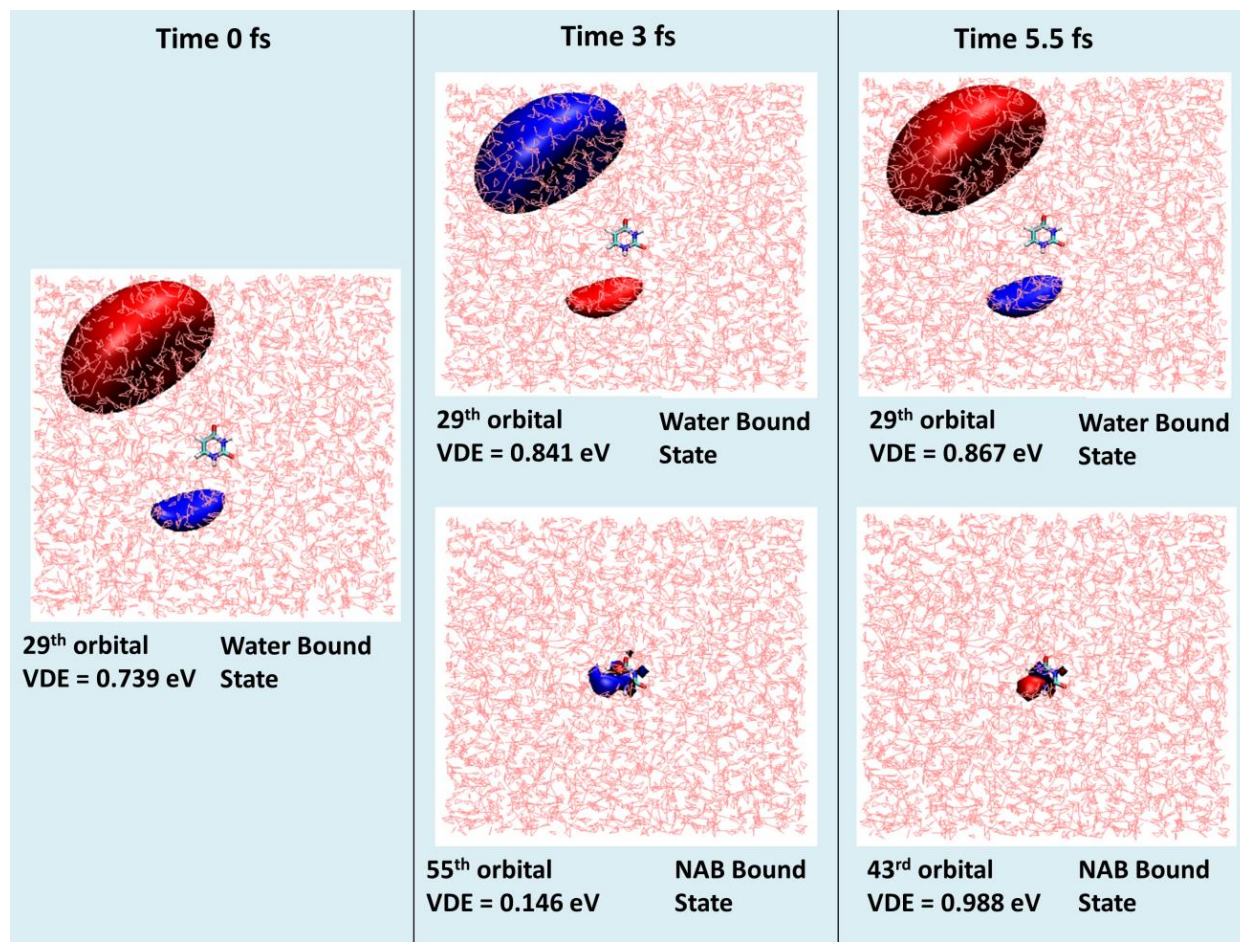

*Figure 4: EA-EOM-DLPNO-CCSD natural orbitals depicting time evolution of anionic state formed by attachment of bulk solvated electron to uracil*